\documentclass[twocolumn,showpacs,superscriptaddress,preprintnumbers,amsmath,amssymb,prb,aps]{revtex4-1}
\usepackage{graphicx}
\usepackage{dcolumn}
\usepackage{bm}

\begin{document}

\title{Observation of three-particle complexes in WS$_2$ monolayers}

\author{J.~Jadczak}\email{joanna.jadczak@pwr.edu.pl}
\affiliation{Department of Experimental Physics, Wroc{\l}aw University of Technology, Wroc{\l}aw, Poland}
\author{J.~Kutrowska-Girzycka}
\affiliation{Department of Experimental Physics, Wroc{\l}aw University of Technology, Wroc{\l}aw, Poland}
\author{P.~Kapu\'sci\'nski}
\affiliation{Department of Experimental Physics, Wroc{\l}aw University of Technology, Wroc{\l}aw, Poland}
\author{J.~Debus}
\author{D.~Kudlacik}
\author{D.~Schmidt}
\affiliation{Experimentelle Physik 2, Technische Universit\"at Dortmund, 44227 Dortmund, Germany}
\author{Y.~S.~Huang}
\affiliation{Department of Electronic Engineering, National Taiwan University of Science and Technology, Taipei 106, Taiwan}
\author{M.~Bayer}
\affiliation{Experimentelle Physik 2, Technische Universit\"at Dortmund, 44227 Dortmund, Germany}
\author{L.~Bryja}
\affiliation{Department of Experimental Physics, Wroc{\l}aw University of Technology, Wroc{\l}aw, Poland}

\begin{abstract}
Atomically thin semiconducting transition-metal dichalcogenides provide novel insights into the physics of many-body effects mediated by Coulomb interactions. Here, we report on temperature-dependent ($T = 7-295$~K) reflectance contrast and photoluminescence studies of three-particle complexes in n-doped WS$_2$ monolayers. In low-temperature reflectance contrast spectra we observe distinct resonances of the neutral exciton, negative trion and exciton bound to a donor ($X$, $X^-$ and $X^D$). For temperatures above 80~K, reflectance contrast signatures of the $X^D$ disappear, whereas the $X^-$ remains detectable up to 240~K, despite the fact that the $X^D$ signal is more red-shifted from the neutral exciton than that of the $X^-$. This experimental observation underlines that in WS$_2$ the dissociation energy of $X^-$ considerably exceeds (factor of 2.5) that of $X^D$. In the laser-power dependent photoluminescence experiments, performed at room temperature, we demonstrate the control of the intensity ratio and energy position of the $X$ and $X^-$ lines, which allows us to evaluate the trion binding energy. Moreover, we demonstrate that the room-temperature PL is sensitive to the environmental gas (ambient, N$_2$, He).
\end{abstract}

\maketitle

Semiconducting transition-metal dichalcogenides (TMDCs), such as MoS$_2$, MoSe$_2$, WS$_2$ and WSe$_2$ have recently attracted a significant attention due to the underlying physics\cite{Mak2010, Splendiani2010, Eda2011, Zhang2014, Cao2012, Sallen2012, Xu2014, Xiao2012, Dery2015, Kormanyos2015, Scrace2015} and the promising optoelectronic applications\cite{Kioseoglou2012, Mak2012, Ross2013, Wang2012, Zhang2014a}, which have been revealed for reducing the thickness to a monolayer (ML). These MLs exhibit common features such as direct band gaps with energies in the visible region at the binary indexed corners $K_{\pm±}$ of the two-dimensional (2D) hexagonal Brillouin zone and the coupling of the spin and valley degrees of freedom. The confinement to a single layer and reduced dielectric screening lead to strong many-body effects mediated by Coulomb interactions. In TMDC monolayers neutral excitons, bound electron-hole pairs ($X=e+h$), exhibit very high binding energies of a few hundreds of meV\cite{Ye2014, He2014, Chernikov2014} leading to their stability at room temperature. Besides neutral excitons the observation of charged excitons in optical spectra of TMDC MLs have been reported\cite{Ross2013, Withers2015, Wang2014, Arora2015, You2015, Wang2015, Mitioglu2013, Chernikov2015, Mak2013, Plechinger2015, Radisavljevic2011, Shang2015, Zhu2015, Peimyoo2014, Mouri2013}. These three-body complexes, also called trions, consist of two electrons and one hole ($X^-$=$2e+h$) or two holes and one electron ($X^+$=$e+2h$). In low temperature photoluminescence (PL) spectra of MoSe$_2$ and WSe$_2$ MLs the exciton and trion transitions have been well resolved\cite{Withers2015, Wang2014, Arora2015, You2015, Wang2015}. However, the emission spectra of the MoS$_2$ and WS$_2$ MLs demonstrate different characteristics\cite{Mitioglu2013}, and the observation of the exciton fine-structure demands special conditions, e.g., the application of a gating or alternatively the selection of samples with appropriate doping levels\cite{Chernikov2015, Mak2013, Plechinger2015}. The differences in the PL spectra between selenide and sulfide MLs are related to the substantial difference of their intrinsic 2D carrier concentration (measured in vacuum), which is about two orders of magnitude higher in sulfides than in selenides\cite{Ross2013, Radisavljevic2011}. 

Despite the intensive studies there are still controversies about the assignment of the $X$ and $X^-$ emission lines in the PL spectra of the sulfide MLs. Nevertheless, in low-temperature reflectance contrast (RC) spectra of WS$_2$, $X$ and $X^-$ resonances have been observed\cite{Chernikov2014}. Moreover, the negative trion binding energy evaluated by different authors from different spectroscopy experiments varies from 15~meV to 38~meV in WS$_2$\cite{Chernikov2014, Mitioglu2013, Plechinger2015, Radisavljevic2011, Shang2015, Zhu2015, Peimyoo2014}, and from 20~meV to 40~meV in MoS$_2$\cite{Mak2013, Mouri2013}. For energies below the $X$ and $X^-$ emission lines, additional features have been shown in low-temperature PL spectra of WS$_2$ and MoS$_2$. They are mainly attributed to excitons bound either to defects or to the surface\cite{Wang2014, Wang2015, Plechinger2015}, but also the observation of biexcitons has been reported in the same energy range of the PL spectra\cite{Plechinger2015, You2015a}. In that context, a comprehensive analysis of temperature-dependent PL and RC spectra has not been performed.

We report on comparative temperature-dependent RC and PL studies of the different exciton complexes in n-type doped WS$_2$ monolayers. Low-temperature RC reveals two strong resonances of the neutral exciton and negative trion and a weak one at an energy that is lower than that of the negative trion, which we attribute to an exciton bound to a donor ($X^D$). For increasing temperature, $X^D$ resonance vanishes, while the $X^-$ resonance is stable up to about 240~K. Our observations confirm recent theoretical calculations\cite{Ganchev2015} indicating that in TMDC monolayers the dissociation energy of trions $X^\pm±$ is much larger than that of excitons bound to a donor or acceptor $X^{D(A)}$. The PL spectra taken at different temperatures are dominated by the emission of localized excitons, while the $X^D$ and $X^-$ features are observed only at low temperatures ($T \leq 10$~K). By tuning the laser power, we show in PL experiments, performed at $T = 295$ K, that we can control the intensity ratio and the energy position of the $X$ and $X^-$ emissions. We thus evaluate the trion binding energy and calculate the intrinsic two-dimensional electron concentration for the WS$_2$ monolayers by determining the energy separation of the $X$ and $X^-$ signatures in the low-temperature RC spectra. Moreover, we demonstrate that the room-temperature PL is sensitive to the environmental gas (ambient, N$_2$, He).

\begin{figure}
\centering
\includegraphics[width=85mm]{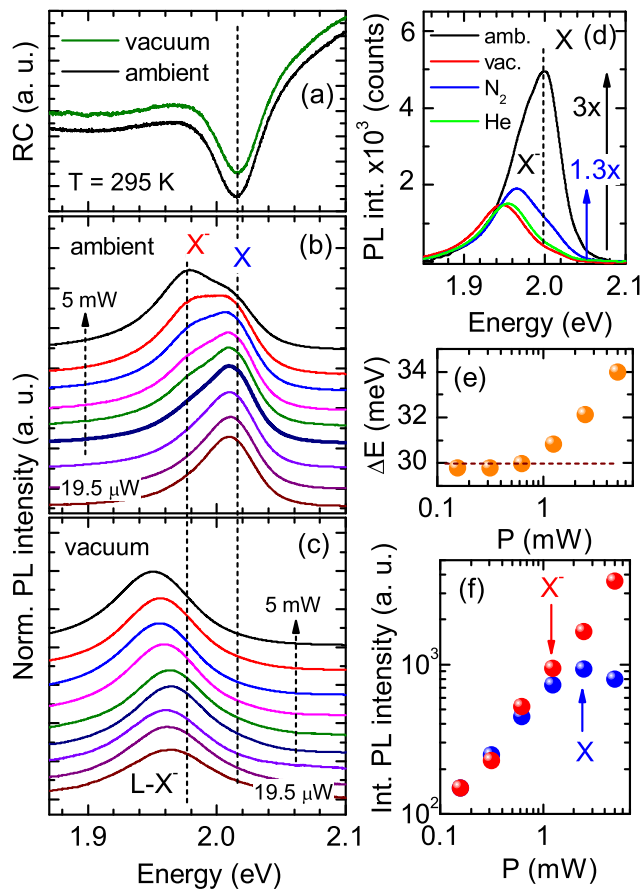}
\caption{Room temperature experiments. (a) RC spectra of a WS$_2$ monolayer measured in ambient and vacuum conditions. Power-dependent evolution of the normalized PL spectra measured in ambient (b) and vacuum (c) conditions. (d) Comparison of PL spectra recorded in different environments; 1-ambient, 2-helium, 3-nitrogen, 4-vaccum. (e) Energy difference between neutral exciton and negative trion peaks as a function of the laser power. (f) Integrated PL intensity of the neutral and charged excitons as a function of the excitation power.}
\label{fig1}
\end{figure}

The studied monolayers of WS$_2$ were prepared by mechanical exfoliation from single crystals on SiO$_2$ (300 nm)/Si substrates. The thicknesses of the mono- and few-layer flakes were established using optical contrast and Raman spectroscopy and subsequently by observing efficient PL. Bulk mixed crystals were grown by the chemical vapor transport technique. Prior to the crystal growth the powdered compounds were prepared from the elements (W:99.99\%  and S: 99.999\%) by reaction at $T=1000^{\circ}$C for 10 days in evacuated quartz ampules. The chemical transport was achieved with Br$_2$ as a transport agent in the amount of about 5 mg/cm$^3$. 

The samples were mounted on the cold finger of a non-vibrating closed-cycle cryostat, where the temperature could be varied from 6 K to about 300 K. Photoluminescence was excited by the $532$~nm line of a diode-pumped solid-state laser. The laser beam was focused on the sample under normal incidence using a $50x$ high-resolution, long-working distance microscope objective ($NA=0.65$). The diameter of the excitation spot was equal to about 1~$\mu$m. The PL emission was collected by the same objective. The spectra were analyzed with a $0.5$-m focal length spectrometer equipped with a $600$ groves/mm grating. A peltier-cooled Si charged-couple device camera was used as detector. The RC measurements were performed with the same setup using a filament lamp as light source.

Figure \ref{fig1}(a) shows room-temperature ($T=295$ K) RC spectra of a WS$_2$ monolayer recorded in vacuum and ambient conditions. One distinct resonance at the energy $E=2018$~meV is detected for both experimental conditions. We attribute this resonance to the neutral exciton ($X$) transition. In Figs.~\ref{fig1}(b) and \ref{fig1}(c) the power-dependent PL spectra measured at $T=295$~K in ambient and vacuum conditions are presented. In the ambient condition, for the lowest laser excitation power, only the $X$ line is observed. When the laser power is increased, an additional line emerges at lower energy and progressively starts to dominate the PL spectra. We assign this line to the negatively charged exciton ($X^-$). The assignments of the $X$ and $X^-$ lines are in excellent agreement with recent reports\cite{Plechinger2015, Zhu2015, Peimyoo2014}. 

The increase in the laser power leads an energetical red-shift of the $X^-$ line, whereas the $X$ line shows a small blue-shift. This effect is related to a photo-induced change of the 2D electron concentration\cite{Tongay2013, Miller2015}. The sulfides, such as MoS$_2$ and WS$_2$, are mostly naturally n-doped\cite{Ross2013}. The 2D carrier concentrations, measured in vacuum, are in the order of $10^{12}-10^{13}$ cm$^{-2}$. In ambient condition, physical adsorption of O$_2$ and H$_2$O molecules effectively depletes the carrier concentration via charge transfer. However, O$_2$ and H$_2$O molecules can gradually be removed by exposure to light\cite{Tongay2013, Miller2015}. In the recent report by Miller et al.\cite{Miller2015} it was outlined that in a MoS$_2$ ML the photo-induced carrier concentration could be changed by almost two orders of magnitude. The same effect is anticipated for WS$_2$. For a low carrier concentration, the energy separation of the $X$ and $X^-$ lines is equal to the trion binding energy. However, for a high 2D electron concentration, during an optical electron-hole transition in the $X^-$ complex an additional electron is excited over the Fermi energy due to the space-filling effect. Accordingly, the relation between the exciton and trion energy positions can be given by the following equation\cite{Mak2013}:
\begin{equation} \label{eq1}
E(X)-E(X^-)=E_{\text{b},X^-}-E_{\text{F}} ,
\end{equation}
where $E(X)$ and $E(X^-)$ are the exciton and trion energy positions in the optical spectra, $E_{\text{b},X^-}$ is trion binding energy and $E_{\text{F}}$ the Fermi energy. As it can be seen in Fig.~\ref{fig1}(e), the dependence of the exciton-trion peak separation on the laser power follows the Eq.~(\ref{eq1}). The minimum line separation of 30~meV estimates the trion binding energy. In Fig.~\ref{fig1}(f) the integrated $X$ and $X^-$ PL intensities are presetend as a function of the laser power. The $X^-$ PL intensity increases linearly with the laser power, whereas the $X$ intensity increases linearly only for low excitation powers and saturates for high ones, when the separation of the $X$ and $X^-$ lines increases above 30~meV. 

In the room-temperature PL spectra, performed in vacuum, we observe one broad line positioned at lower energy with respect to the $X$ and $X^-$ lines that are detected in ambient condition. This line experiences a red-shift in energy for rising laser power. We attribute it to the superposition of radiative recombination of the trions and localized excitons ($L$). Further details on this transition are presented within the frame of the temperature-dependent PL measurements. In Fig.~\ref{fig1}(d) we compare the room-temperature PL spectra measured in ambient, N$_2$, He,and vacuum conditions. The PL spectra recorded in vacuum and under He gas are almost identical, whereas in the N$_2$ environment the PL lines are slightly shifted to high energies. This observation is in contrast to the report of S.~Tongay et al.~\cite{Tongay2013}, where similar PL spectra for a MoS$_2$ ML were detected in vacuum and N$_2$ environment. Also, they have detected an about 100 times higher PL intensity for the MoS$_2$ ML in ambient than in vacuum conditions, while in our PL studies, for WS$_2$ monolayers, we observe that the signal in ambient condition is only about 3 times higher than in vacuum.

\begin{figure}
\centering
\includegraphics[width=85mm]{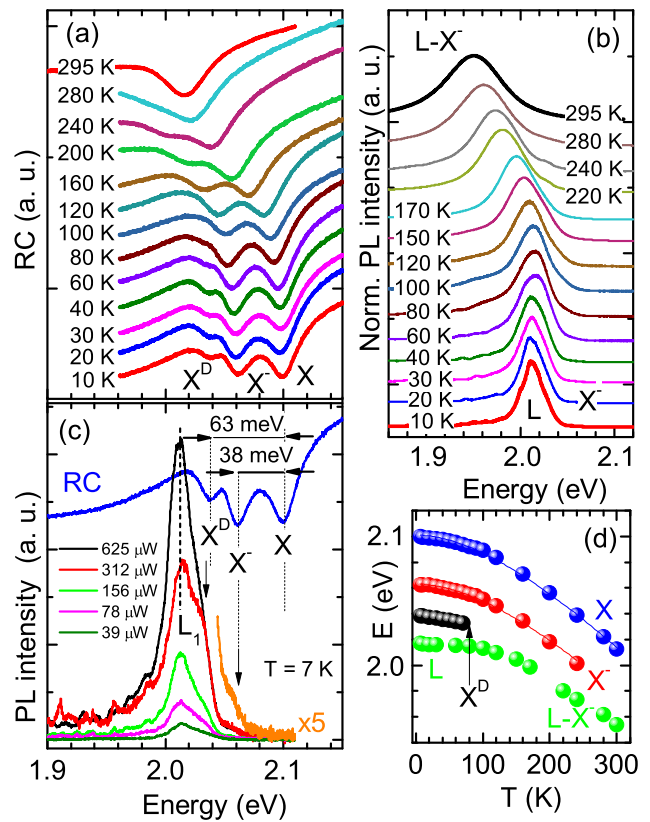}
\caption{(a) Comparison of the PL and RC spectra of a monolayer WS$_2$ at $T=7$ K. (b) Temperature evolution of the RC (b) and PL (c) spectra. (d) Energy position of $X$, $X^-$ and $X^D$ as a function of the temperature. The solid lines represent the numerical fits to the experimental data following Eq.~(\ref{eq2}).}
\label{fig2}
\end{figure}

The comparative room temperature PL and RC measurements, performed in ambient and vacuum environments, provide a thorough understanding of the temperature-dependent experiments performed in vacuum. The temperature-dependent measurements can also be performed in gaseos or liquid He environments, but, as we have shown, the PL and RC spectra in vacuum and He atmosphere are similar to each other; see Fig.~\ref{fig1}(d). It results from the fact that He is not physisorbed on the surface of the WS$_2$ monolayer; hence, it does not change the 2D carrier concentration. In Fig.~\ref{fig2}(a) the temperature evolution the RC spectra is presented. At high temperatures one resonance of the neutral exciton can be seen in the spectra. When the temperature is decreased, two new resonances appear, one at $T<240$ K and an other at $T<80$ K with energies of 38~meV and, respectively, 63~meV in relation to the $X$ energy. We assign these resonances to the optical transitions of negative trions (low-energy resonance) and to that of excitons bound to a donor ($X^D = D^{+} + e + h$, high-energy resonance). Our assignment of $X^D$ is based on the recent theoretical report of Ganchev et al.~\cite{Ganchev2015}. In numerical calculations of various three-particle complexes in 2D semiconductors they have found out that in TMDC monolayers the dissociation energies of trions are much larger than that of excitons bound to a donor or acceptor, so the $X^\pm$ are more resilient to heating, despite the fact that they are positioned at higher energies in optical spectra than $X^{D(A)}$. For WS$_2$, assuming the electron-hole mass ratio to $m_e/m_h\approx≈0.6$, the calculated dissociation energy of $X^-$ is about 2.5 times higher than that of $X^D$. This value is in excellent agreement with our experimental observations, since $X^D$ disappears in the RC spectra at an about 3 times lower temperature than $X^-$. 

The temperature shift of the $X$ and $X^-$ resonances follows an empirical Varshni formula\cite{Varshni1966}, describing the temperature dependence of the energy gap in many semiconductors:
\begin{equation} \label{eq2}
E_{\text{g}}(T)=E_{\text{g}}(0)-\alpha T^2/(T+\beta) ,
\end{equation}
where $E_{\text{g}}(0)$ is the energy gap at 0 K, and $\alpha$ and $\beta$ are fitting parameters. Since the energy separation of the $X$ and $X^-$ resonances is temperature independent, we use the same $\alpha$ and $\beta$ parameters. The fits to our experimental data yield the values $\alpha=6.3\times10^{-4}$~eV/K, $\beta=350$~K, and $E_{\text{g}}(0)=2099$~meV for the neutral exciton, and $E_{\text{g}}(0)=2061$~meV for the negative trion. The parameters are comparable to those obtained in previous studies on $X$ and $X^-$ complexes for WS$_2$ MLs\cite{Plechinger2015}. From the energy separation of the $X$ and $X^-$ resonances and Eq.~(\ref{eq1}) we evaluate the Fermi energy to $E_{\text{F}}=8$~meV. Using the equation $n=m_eE_{\text{F}}/\pi\hbar^2$ and an electron effective mass of $m_e=0.37$,\cite{Xiao2012} we calculate the intrinsic 2D electron concentration to $n=1.24\times10^{12}$/cm$^2$.

Finally, let us analyze the temperature evolution of the PL spectra presented in Fig.~\ref{fig2}(b). In these experiments the laser power was kept relatively low at 100 $\mu$W. In Fig.~\ref{fig2}(c) the PL and RC spectra measured at 7 K are compared. The PL spectra are shown for different laser power excitations. At $T=7$~K the PL spectrum consists of two features: (i) a weak negatively charged exciton line and (ii) a broad feature at low energy, whose width may hint at a localized exciton bound to defects or the surface\cite{Wang2014, Wang2015, Plechinger2015}. Within this broad feature we can identify at least two relatively broad lines located at high energies, and a series of relatively narrow lines at low energies. In comparison with the RC spectra we attribute the highest-energy line within the broad feature to the radiative recombination of the exciton bound to the donor. Our assignment of the $X^-$ and $X^D$ lines in the PL spectra is based on their similar shift to lower energies with respect to their positions in the RC spectra. The origin of the low-energy PL lines (with regard to $X^D$) is still under discussion. In the study of Plechinger et al.~\cite{Plechinger2015} it was reported that the line, which we denominate as $L_1$, may be a superposition of defect-bound exciton and biexciton ($XX$) emissions. They observed a non-linear increase of the $L_1/XX$ photoluminescence intensity as a function of the laser power density, whereas the $X$ and $X^-$ PL intensities exhibited a linear shift. In our power-dependent PL measurements performed at $T=7$ K we observe a similar non-linear increase of the PL intensity for the $L_1$ line. However, we do not detect well-resolved $X$ and $X^-$ PL lines to do comparisons. The series of narrow PL lines at low energies, observed also in other TMDCs, were attributed to surface-bound exciton emissions, being promising as single photon emitters\cite{Koperski2015}. The temperature shift of the broad low-energy PL feature ($L$) differs strongly from that of the $X$ and $X^-$ resonances, detected in the RC spectra. Its energy position is essentially not affected by the growth of temperature from 7~K to about 120~K, but exhibits a red-shift with a further increase in temperature. Its corresponding rate is much lower than that of the $X$ and $X^-$ resonances, see Fig.~\ref{fig2}(d). This behavior of the PL spectra is likely related to a shift of the spectral weight from localized excitons to trions for rising temperature.

In conclusion, we report on temperature-dependent RC and PL studies of three-particle complexes in n-doped WS$_2$ monolayers. In low-temperature RC spectra we observe three distinct resonances that are attributed to the neutral exciton, negative trion, and exciton bound to a donor. A temperature increase results in the disappearance of the $X^D$ resonance at about 80~K, whereas the $X^-$ resonance remains visible up to temperatures of 240~K, although the $X^D$ complex is more strongly red-shifted from the neutral exciton than the $X^-$ complex. The dissociation energy of the negative trion is shown to be larger than that of a neutral exciton bound to a donor. In comparative PL and RC experiments performed at different experimental conditions we estimate the trion binding energy and the intrinsic two-dimensional electron concentration. Additionally, we demonstrate a significant sensitivity of the room-temperature PL emission on the gaseous environment, which may be used in gas sensing applications. 

This work was partly supported by the Polish NCN Grant No. 2013/09/B/ST3/02528 and the Polish-Taiwanese Joint Research OSTMED.

\end{document}